\documentclass[conference]{IEEEtran}

\usepackage{amsthm}
\usepackage[cmex10]{amsmath}
\usepackage{amsfonts}
\usepackage{amssymb}

\ifCLASSINFOpdf
\usepackage[pdftex]{graphicx}
\usepackage{subfigure}
\else

\usepackage{graphicx}
\usepackage{subfigure}
\fi
\usepackage{epstopdf}
\usepackage{color}

\usepackage[font=normalsize]{caption}

\usepackage{algorithmic,algorithm}

\usepackage[numbers,sort&compress]{natbib}
\usepackage{color}

\usepackage{url}

\usepackage{mdwtab}

\begin{document}
	
	\title{A Fog-based Architecture and Programming Model for IoT Applications in the Smart Grid}


\author{\IEEEauthorblockN{Pan Wang\IEEEauthorrefmark{1}, Shidong Liu\IEEEauthorrefmark{2},
		Feng Ye\IEEEauthorrefmark{3} and Xuejiao Chen\IEEEauthorrefmark{4}}
		\IEEEauthorblockA{
		\IEEEauthorrefmark{1}Nanjing University of Posts \& Telecommunications, Nanjing, China\\
		\IEEEauthorrefmark{2}Global Energy Interconnection Research Institute, Nanjing, China\\
		\IEEEauthorrefmark{3}Department of Electrical \& Computer Engineering, University of Dayton, Dayton, OH, USA\\
		\IEEEauthorrefmark{4}Department of Communication, Nanjing College of information Technology, Nanjing, China}}	
\maketitle
	
\begin{abstract}
The smart grid utilizes many Internet of Things (IoT) applications to support its intelligent grid monitoring and control. The requirements of the IoT applications vary due to different tasks in the smart grid. In this paper, we propose a new computing paradigm to offer location-aware, latency-sensitive monitoring and intelligent control for IoT applications in the smart grid. In particular, a new fog-based architecture and programming model is designed. Fog computing extends computing to the edge of a network, which has a perfect match to IoT applications. However, existing schemes can hardly satisfy the distributed coordination within fog computing nodes in the smart grid. In the proposed model, we introduce a new distributed fog computing coordinator, which periodically gathers information of fog computing nodes, e.g., remaining resources, tasks, etc. Moreover, the fog computing coordinator also manages jobs so that all computing nodes can collaborate on complex tasks. In addition, we construct a working prototype of intelligent electric vehicle service to evaluate the proposed model. Experiment results are also presented to demonstrate that our proposed model exceed the traditional fog computing schemes for IoT applications in the smart grid. 
\end{abstract}

\section{Introduction}\label{sec:intro} 

Many Internet of Things (IoT) applications will be deployed in the smart grid to enable efficient operation of the grid~\cite{Yan_2013,Ye_2015}. The IoT applications in the smart grid could be used for monitoring the power transmission line, the substation, managing electric vehicle charging/discharging, user information collection etc. Due to the increasing number of IoT nodes in the smart grid, e.g., smart meters, phasor measurement units, etc., the traditional cloud data center computing paradigm is not able to meet the requirements of IoT applications in the smart grid. The smart grid requires the IoT applications to have high bandwidth, low latency and location-awareness. To tackle those issues, we propose a fog computing based architecture and programming model for IoT applications in the smart grid. Fog computing was proposed by Cisco to extend the cloud computing paradigm to run distributed applications~\cite{bonomi2012}. As part of cloud computing, fog computing not only deals with latency-sensitive applications at the edge of a network, but also deals with latency-tolerant tasks with more power computing nodes in the middle of a network. In the upper layer of a fog, cloud computing supported by powerful data centers can be applied for further processing.


Many research work has been conducted in the area of fog computing in the recent years. For example, the authors of~\cite{Aazam2014gateway} proposed a fog computing and smart gateway based communication design for cloud of things. The authors of~\cite{Faruque2016energy} brought forward an energy management platform based on the fog computing architecture. The authors of~\cite{Nazmudeen2016PLC} introduced a distributed processing framework for data aggregation based on fog computing architecture. The authors of~\cite{Okay2016smartgrid} presented a fog computing based smart grid model, which comprises smart grid layer, fog layer and the cloud layer. Based on the concept of fog computing, the authors of~\cite{Yan2016AMI} proposed a portable data storage and processing solution to advanced metering infrastructure. However, the existing fog-based computing architecture hardly mention distribution coordination within fog computing nodes. Since the IoT nodes in the smart grid are usually closely related, our proposed fog-based architecture introduces a fog computing coordinator to better coordinate fog nodes. In particular, the coordinator gathers information of fog computing nodes in the same area periodically. In addition, the coordinator is also in charge of jog assignment to fog computing nodes to collaborate on complex tasks.

In addition, research work on programming model for IoT application has also been carried out in recent years. For example, the authors of~\cite{Satoh2013} proposed a framework based on MapReduce for data processing at the edges using mobile agents. The authors of~\cite{Cherrier2011} introduced a distributed logic for IoT services based on OSGi to improve modularization programming. The authors~\cite{Hong2013} brought forward a programming model using mobile fog for large scale applications on IoT. However, existing schemes mentioned before cannot meet the new requirements of IoT application in smart grid, especially the distributed coordination within fog computing nodes. Therefore, we propose a programming model specifically for the proposed fog-based architecture in the smart grid.

\begin{figure*}[ht!]
	\centering\includegraphics[width=6.2 in]{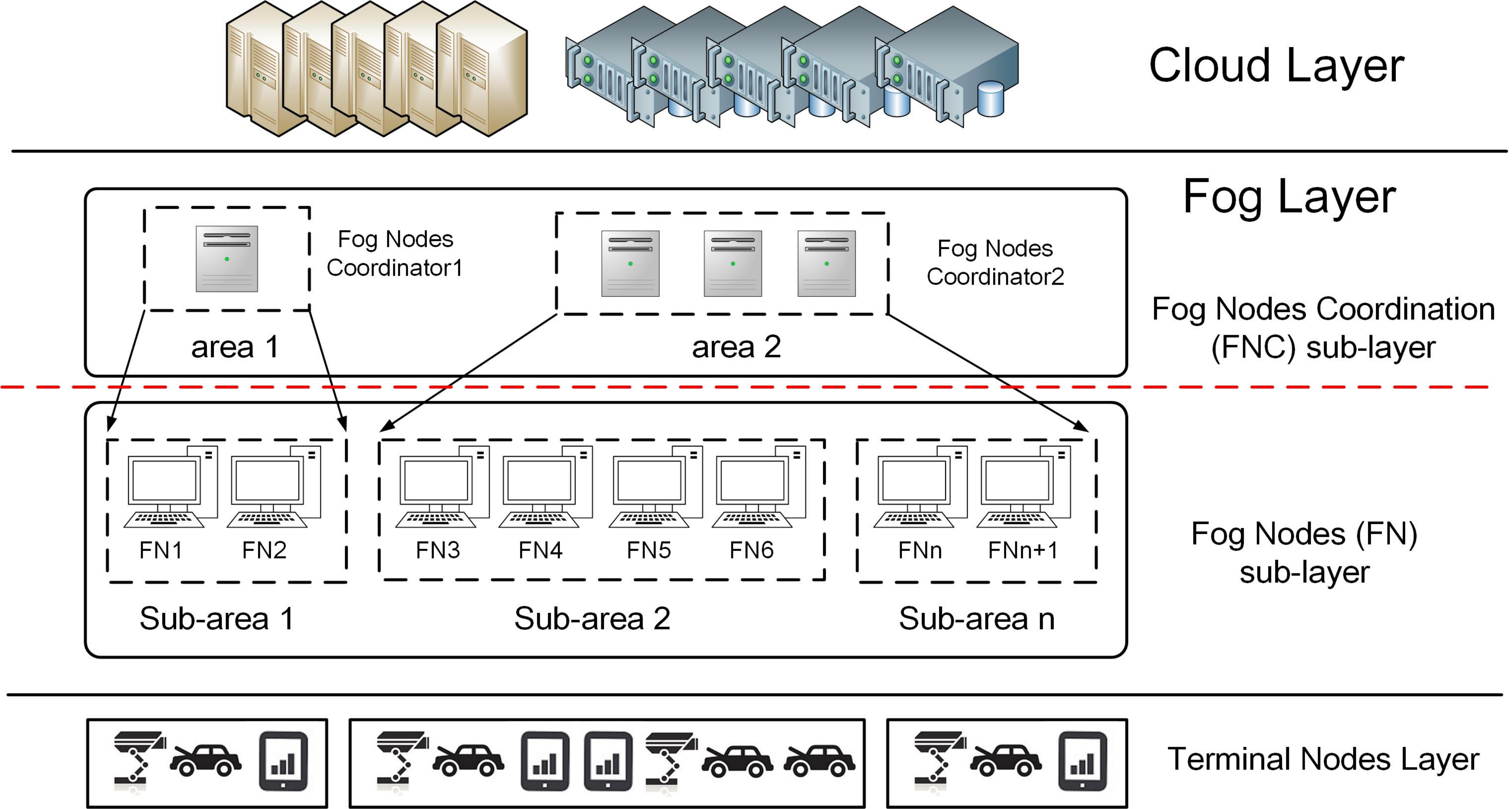} 
	\caption{Fog Computing Based Architecture for IoT application in Smart Grid}\label{fig:Fig_1_architecture} 
\end{figure*}

In summary, the contributions in this paper are mainly twofolds. Firstly, we propose a new distributed fog-based computing architecture for IoT applications in the smart grid. To improve the performance in latency, we integrate a fog computing coordinator in the proposed architecture. Secondly, we propose a programming model that specifically focuses on the proposed architecture. The remaining of the paper is organized as follows. The new Fog Computing architecture is presented in Section~\ref{sec:archi}. The programming model corresponding to the architecture is discussed in Section~\ref{sec:program_model}. Evaluation and experimental results are presented in Section~\ref{sec:eval} to demonstrate our proposed architecture as well as programming model. Finally, the conclusion is drawn in Section~\ref{sec:conclusion}.

\section{Fog-based Architecture for IoT applications in the Smart Grid}\label{sec:archi}

In this section we first identify several requirements that need to be considered for IoT applications in the smart grid. We then give the proposed fog-based architecture.

\subsection{Requirements for IoT applications in the Smart Grid}

The requirements for IoT applications in the smart grid mainly include latency, distributed coordination, location awareness and mobility support. 

\begin{itemize}
	\item \textbf{Latency sensitivity}: Many IoT applications in the smart grid depend on real-time decisions~\cite{Chen2017}. For example, substations in the smart grid are equipped with various sensors to monitor status of power transmission. In this scenario, a longer latency may lead to serious accident such as power failure.

	\item \textbf{Distributed coordination}: There are always a lot of sensors distributed in a certain geographic area, e.g., charging piles for electric vehicles. It is important to cordinate multiple sensors or nodes from several areas to provide services from IoT applications in the smart grid. 
	
	\item \textbf{Locations awareness and mobility support}: The end devices in many IoT applications are mobile in the smart grid, e.g., electric vehicles. It is challenging for fog computing to aggregate and process data close to the source. 
\end{itemize}

\subsection{Proposed Fog-based Architecture}

An overview of the proposed fog-based computing architecture for IoT applications in the smart grid is shown in Fig.~\ref{fig:Fig_1_architecture}. The proposed architecture comprises three layers: terminal layer, fog layer and cloud layer. \emph{Terminal nodes layer} is the bottom layer which consists of smart devices, which are responsible for transmitting sensed data and event logs to the upper layer. \emph{Fog layer} is the middle layer which consists of fog nodes. The fog nodes are deployed at the edge of a network in order to extend the processing ability of a cloud center. \emph{The cloud layer} is the upper most layer in this architecture. This layer consists of powerful servers, such as data centers, which are responsible for analyzing massive historical data.

Our focus in the proposed architecture is on the fog layer. Compared with the traditional fog computing model, our fog layer is divided into fog nodes (FN) sub-layer and fog nodes coordination (FNC) sub-layer. With  computing and storage capability, the fog nodes in the FN sub-layer provide a mechanism for migrating processing logic to the edge of the network. The FN sub-layer also has the aggregation capability for the sensed data from the terminal nodes layer. After being gathered and anlyzed, some of the data is fed back to the active nodes in terminal nodes layer to complete the real-time response and process to the emergency event. The rest of the data is transmitted to the FNC sub-layer.

\begin{figure*}[ht!]
	\centering\includegraphics[width=6.5 in]{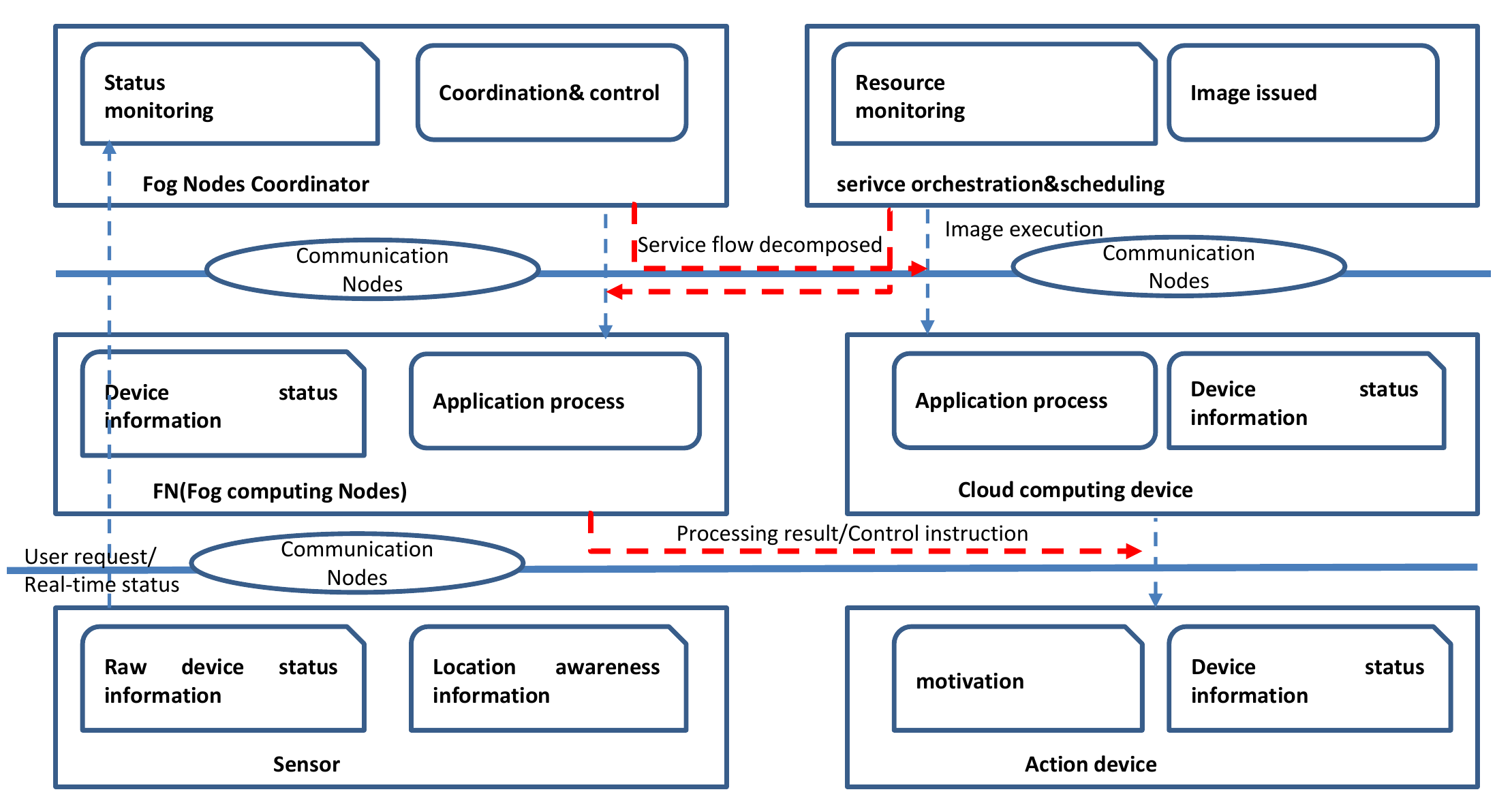} 
	\caption{system model}\label{fig:Fig_2_model} 
\end{figure*}

The FNC sub-layer consists of multiple coordinators that are distributed in service areas. In this layer, fog nodes are divided into several clusters, where some are euipped  with computing and storage capability according to some principle. Such equipment is named as fog computing coordinators (FCN) for simplicity. The FCNs focus on coordinating the fog nodes to deal with some complex tasks. For example, the problem of quering a suitable charging station for moving electric vehicles. In addition to undertaking data analysis and processing of the sub-region, an FCN is also responsible for coordinating all fog computing nodes in the region to further improve application performance by using parallel computing capabilities.

\subsection{Studied System Model}

The studied system model is shown in Fig.~\ref{fig:Fig_2_model}. It consists of sensors, action devices, communication nodes, fog computing nodes, cloud computing servers, FNC, service orchestration and scheduling servers called OSS servers. Note that the FNS sub-layer is composed of FNCs from all service areas. An FNC receives requests from terminal nodes and decomposes the service data flow according to the resource usage and service flow capacity of each FN and cloud computing servers. The jobs are then dispatched to related fog nodes by the FNC. In the end, the FNC collects all the execution results and make the final decisions as instruction to the action devices. Fog nodes in the same layer need to achieve a complete user request with the coordination of an FNC because there is no guarantee of direct connection between any two of them. The interaction between a fog node and the FNC can be achieved using relay communication, for by flow table if using software-defined network controller.

Moreover, we introduce OSS servers in Cloud Computing Center. The OSS servers can decompose services data flow based on resource usage and capacity collected from computing nodes. OSS servers dispatch job execution images to computing nodes in a way that is similar to traditional virtual machine with better security isolation, or docker container with less starting latency. Nonetheless, OSS servers mainly aim at initialization deployment of new applications provided by service providers and setting up service related execution logic on computing nodes. In contrast, an FNC provides application services that can meet the functionality and quality-of-service (QoS) requirements for end users by resource allocation and service logic coordination deployed in various computing nodes after receiving service requests from end users.

\begin{figure*}[ht]
	\centering\includegraphics[width=6.in]{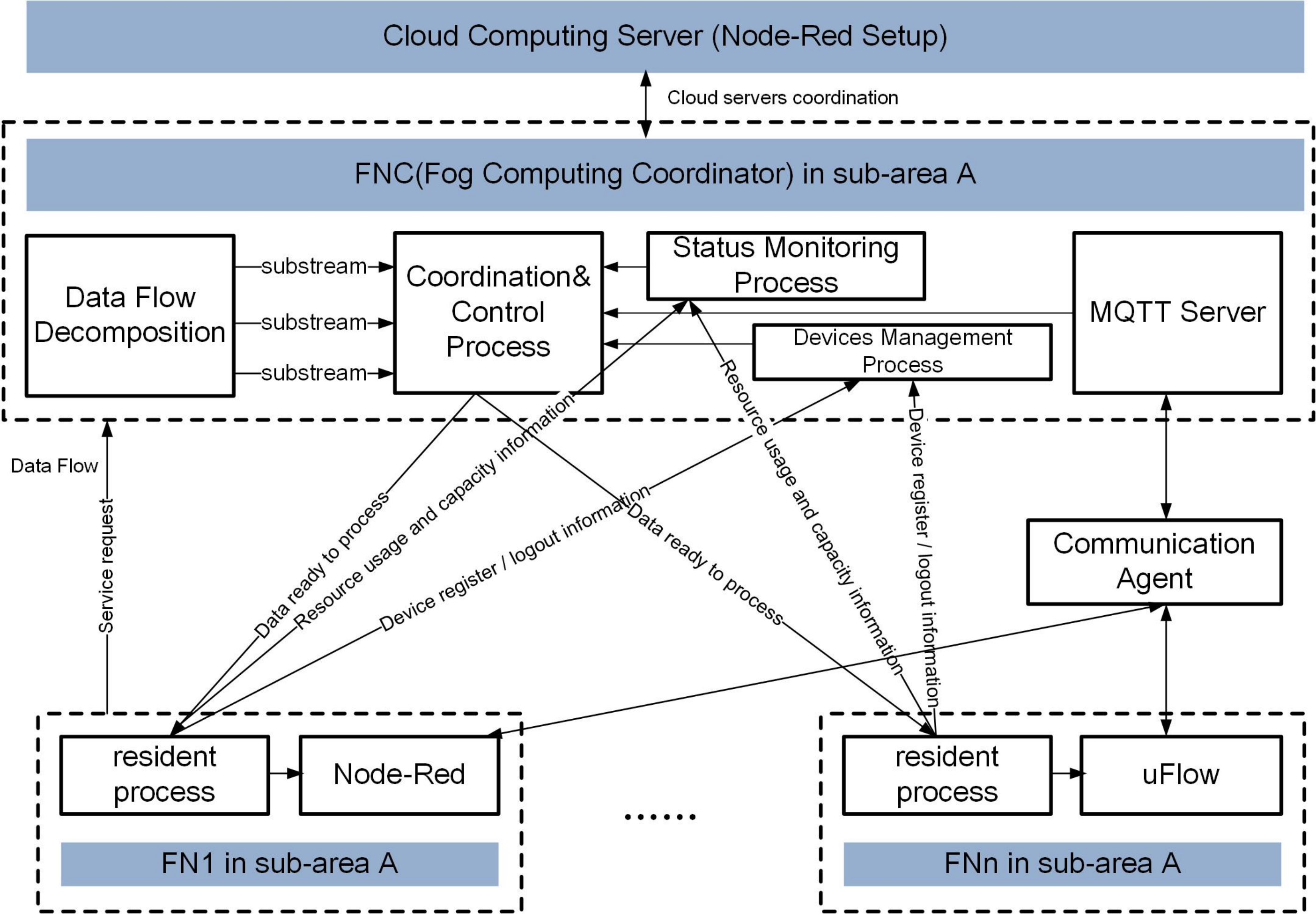} 
	\caption{Proposed distributed coordination dataflow programming model for the fog-based architecture.}\label{fig:Fig_3_programmingmodel} 
\end{figure*}

\section{Distributed Coordination Dataflow Programming Model}\label{sec:program_model}

Many of those IoT applications are distributed systems in the smart grid. A typical programming framework for such systems is WoTKit processor based on WoTKit platform~\cite{Blackstock2012} and NR of IBM~\cite{ibm}. WoTKit is developed on JAVA Spring framework, where developers can run data flow programs by creating links among modules. However, WoTKit is designed for deployment on server level. It should be better to use NR framework in IoT application. The NR framework was designed for application-level development in a single computing unit, including nodes of input, output, processing and visual developing environment based on Web. Although some features towards distributed computing has been added to the NR framework~\cite{Giang2015}, the traditional programming model based on \emph{request-response} cannot achieve real-time processing for the smart grid. Therefore, we propose a new programming model which is based on data flow programming~\cite{Johnston2004}.

An overview of the proposed distributed coordination dataflow programming model is shown in Fig.~\ref{fig:Fig_3_programmingmodel}. With the control of FNCs, cloud servers and fog nodes distributed in different geographic areas can perform data analysis and process of application services together. We assume that there are two types of computing nodes: one has rich computing resource, which uses Node-Red as distributed data flow computing framework; the other has limited resource, which uses uFlow as flow processing framework.

\begin{figure}[ht!]
	\centering
	\fbox{\begin{minipage}{0.45\textwidth}
			\noindent\textbf{Pseudocode of the migration process}: 
			
			1：  procedure migration\_source(nodeID)
			
			2:   	obtain the actual latency T to node whose ID is nodeID
			
			3:   	compare T with threshold Tupper
			
			4:   	if  T  less than  Tupper
			
			5:      	return;
			
			6:   	else 
			
			7:      	obtain the best candidate Vb from candidate group V
			
			8:      	V = V - Vb
			
			9:      	send a Start Migration message to node Vb
			
			10:     	wait for response Resp
			
			11:     if  Resp is ACCEPT then
			
			12:        S = on\_migration\_start(nodeID)
			
			13:        send Object State message to node Vb
			
			14:        release local resource of dataflow relate to nodeID
			
			15:        return
			
			16:     else 
			
			17:        if V is empty
			
			18:          warn the message(can not migrate)
			
			19:          return
			
			20:        endif
			
			21:     endif
			
			22:     goto 7
			
			23:  endif
			
			24: end procedure
		\end{minipage}
	}
\end{figure}

The resident processes in every distributed computing node are are responsible for collecting information such as resource and capacity. The collected information is then reported to the upper layer so that the FNC can make better decision and instruction. After receiving substream and data, a fog node translates the data flow into instructions that can be identified and executed at terminal nodes. Take the example of electric vehicle parking services. When a vehicle under the control of a sub-area fog node applies for a parking service, the FNC of this sub-area will dispatch the request to all fog nodes within this sub-area. After distributed coordinative processing by all fog nodes, the FNC can provide the best parking information for the requester. After processing by the fog nodes locally, the FNC should report all the statistical data to cloud servers for future analysis.

A challenge exists due to the mobility of terminal nodes. The final data flow made by the FNC may not fit the current network condition. In this case, fog nodes need coordinate by themselves to fix this situation, which is defined as fog computing nodes migration. According to the initiator of migration, we can divide the migration into two types: one is initiated by FN; the other is initiated by terminal nodes. Two APIs used during the process of migration are defined as follows: 
\begin{itemize}
	\item State on\_migration\_start (nodeID): Invoked by migration device before migration process start. It returns a state object for flow information waiting for migration.
	
	\item Void on\_migration\_end (state\_s): Invoked by migration target device after receiving request message from migration initiator.
\end{itemize}

A fog nodes will decide whether to migrate or not by comparing the QoS of terminal nodes and a predefined threshold at the fog node during the process of sub-stream computing. This process is illustrated in lines 2-5 in the given pseudocode. Once decided, the rest of the process is as follows: 1) choose the best nodes from node group with proper capacity, resource and QoS requirements, see lines 7-8. 2) Send a migration start request message to alternative nodes and wait for response, see lines 9-10. 3) If alternative nodes accept, then call on\_migration\_start to get all status information of current nodes and send to alternative nodes. Meanwhile, free all related resources of data flow ready to migrate, see lines 11-15. 4) If alternative nodes do not accept, send the warning message,like can not migrate, and quit. Otherwise, choose the next alternative node from node group and repeat above-mentioned. Target nodes will call on\_migration\_end to take over following work after receiving migration status message from the migration initiator.

\section{Evaluation and Experimental Results}\label{sec:eval}

\subsection{Evaluation Settings}

We use an electric vehicle intelligent service to evaluate the proposed fog-based architecture and the programming model. The evaluated system is composed of electric vehicles, charging piles, a regional coordinator, a regional application server, a cloud service center, a communication proxy server and basic communication networks. Assume that electric vehicles report real-time information and request for services through sensing devices at the edge of smart grid communications network. The charging pile devices are located at the edge of the network as fog computing devices. Such a fog node processes the data according to the established application logic. It also transmits the processed data through the communication proxy service to the application server of the region or the remote cloud service center according to control of the area coordinator. The regional application server has a cloud service center that provides services to users. The difference is that the regional application server focuses more on providing some geographically related or strict requirements of latency-sensitive services (such as navigation, intelligent parking, etc.). Whereas the cloud center server provides some delay-tolerant analysis and forecasting services. Services areas are set based on densities of charging piles and the sizes of traffic flows. A regional coordinator is set for each area. The coordinator completes the data flow chart and transmits data to corresponding devices based on the service requests from users, as well as remaining resources, capabilities, and location information reported by the equipment. The coordinator also coordinates these devices to complete the related service logic. For services that require cross-domain provisioning, coordination is offered by regional application servers in different regions.

We choose the IBM's NR framework as an implementation tool for application development, and deploy a stream-based micro-runtime environment uFlow over resource-limited IoT devices. In the uFlow environment, we use Lua as the programming language, MQTT as a communication protocol, respectively. In order to provide charging services to the electric vehicles, the system selects the most suitable charging pile to the requesting electric vehicle. The evaluation is conducted using two cases: 

\begin{itemize}
	\item \emph{Using the traditional fog computing architecture.} In this case, electric vehicles directly transmit the requests to all charging piles in a certain range. After the calculation, the charging pile that can meet the requirements returns the confirmation to the electric vehicle. 
	\item \emph{Using the fog computing coordinator architecture.} In this case, electric vehicles will directly send requests to the fog computing coordinator in the area. According to the information of the charging piles, the coordinator will forward the request to some charging piles with the possibility of providing the service. A response is returned to the electric vehicle by the most eligible charging pile.
\end{itemize}

\subsection{Evaluation and Simulation Results}

We mainly evaluate the different of application latency between the traditional fog computing architecture and our proposed architecture based on fog computing coordinator. We used 20 software terminal nodes running on embeded system to emulate electric vehicles; 10 software fog nodes to emulate charging piles; and 2 software nodes to emulate FNC nodes. The range of all entity is defined as a circle with a diameter of 2000 meters.

\begin{figure}[ht!]
	\centering\includegraphics[width=3.6 in]{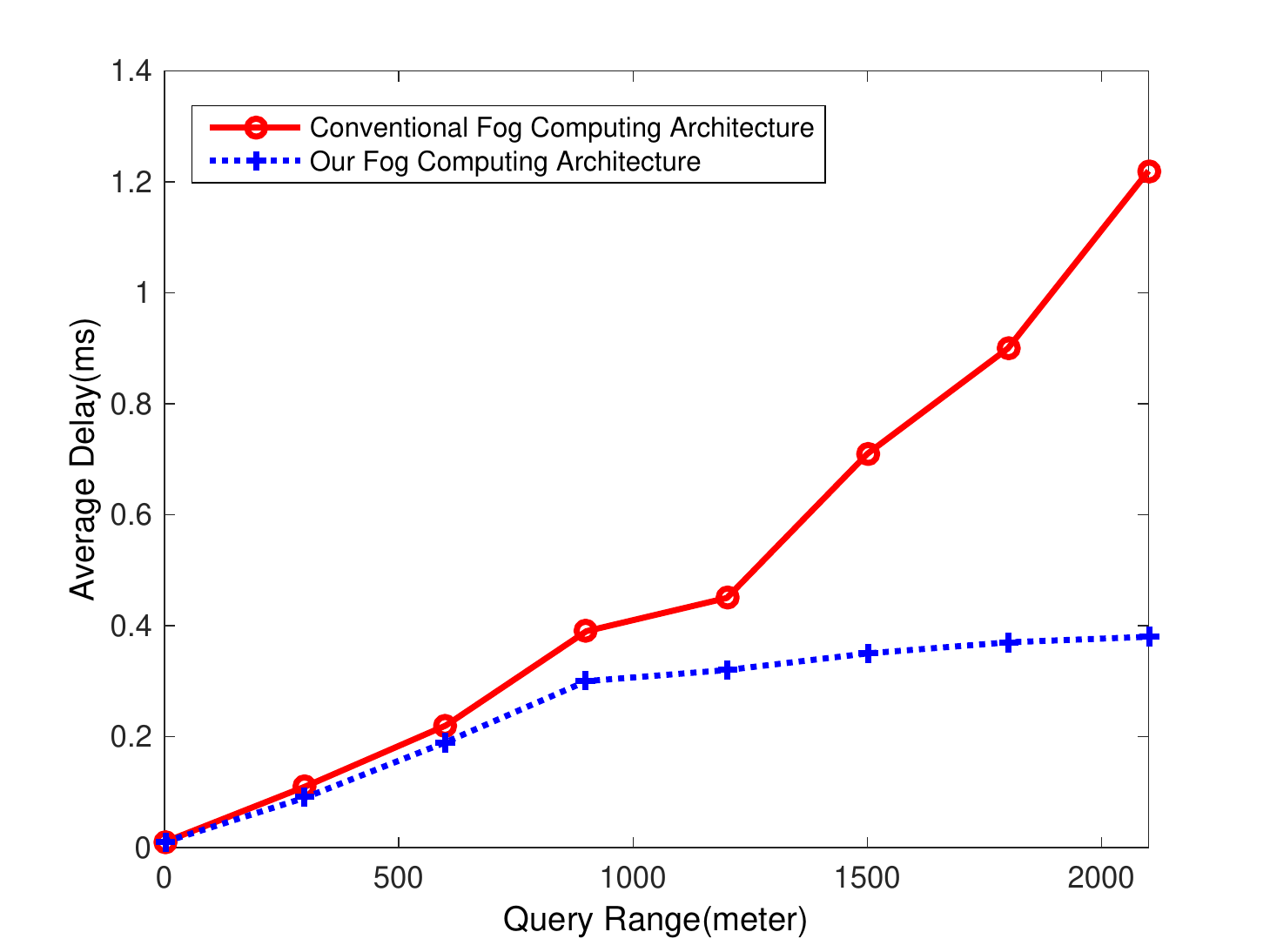} 
	\caption{Query range and application delay.}\label{fig:Fig_4_queryrange} 
\end{figure}

Fig.~\ref{fig:Fig_4_queryrange} shows the relationship between the application latency and the query distance of the two architectures. It can be seen that when the query distance is short, the application latency is similar between the two architectures. Nonetheless, when the query distance is larger, our architecture outperforms the traditional architecture by having a lower application latency.

\begin{figure}[ht!]
	\centering\includegraphics[width=3.6 in]{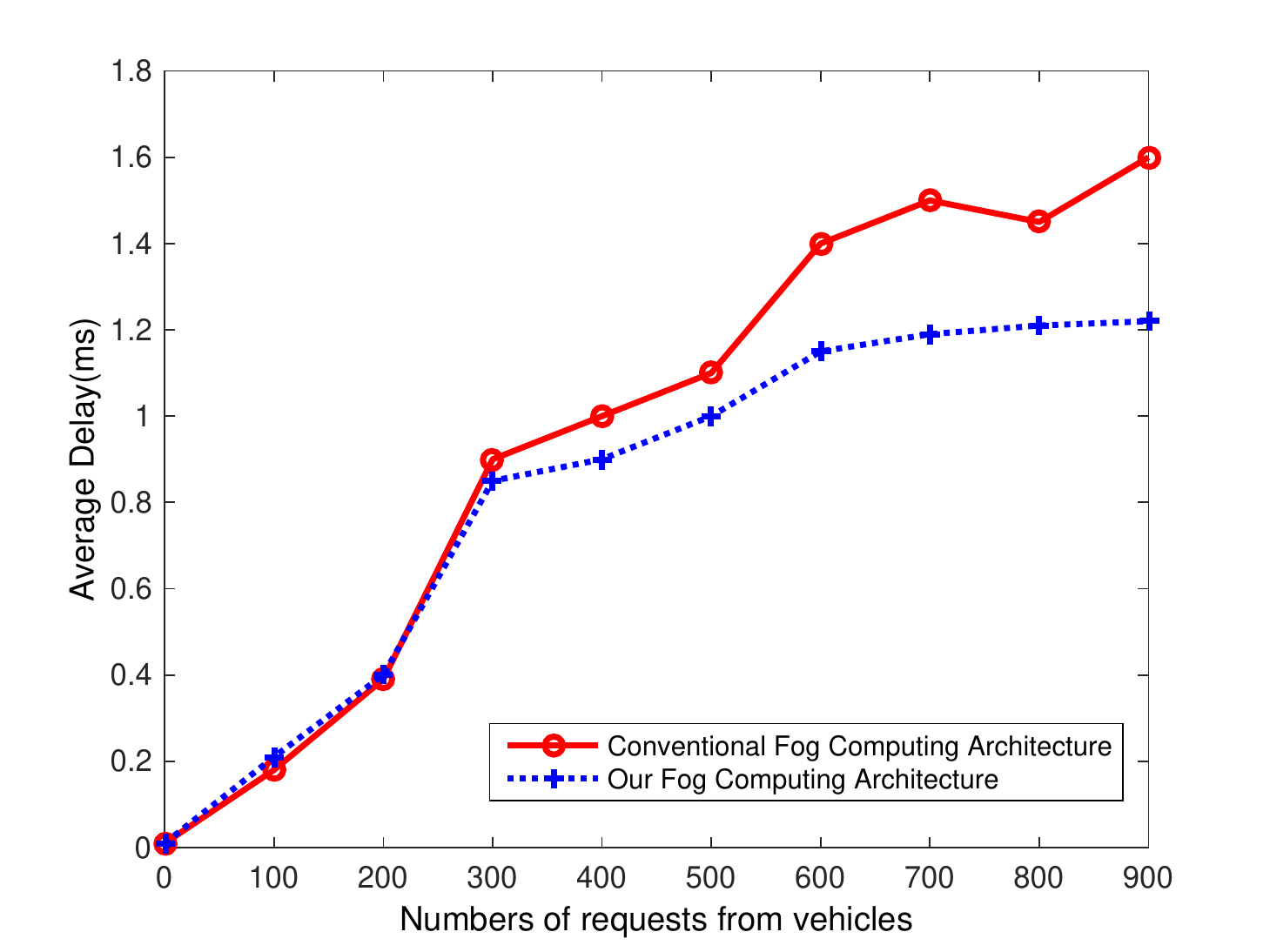} 
	\caption{Service requests and application delay.}\label{fig:Fig_5_requests} 
\end{figure}

Fig.~\ref{fig:Fig_5_requests} shows the relationship between the number of service requests from electric vehicles and application delay. Obviously, as the numbers of service request increases, the proposed architecture has a significantly lower delay compared to the traditional architecture.

\begin{figure}[ht!]
	\centering\includegraphics[width=3.5 in]{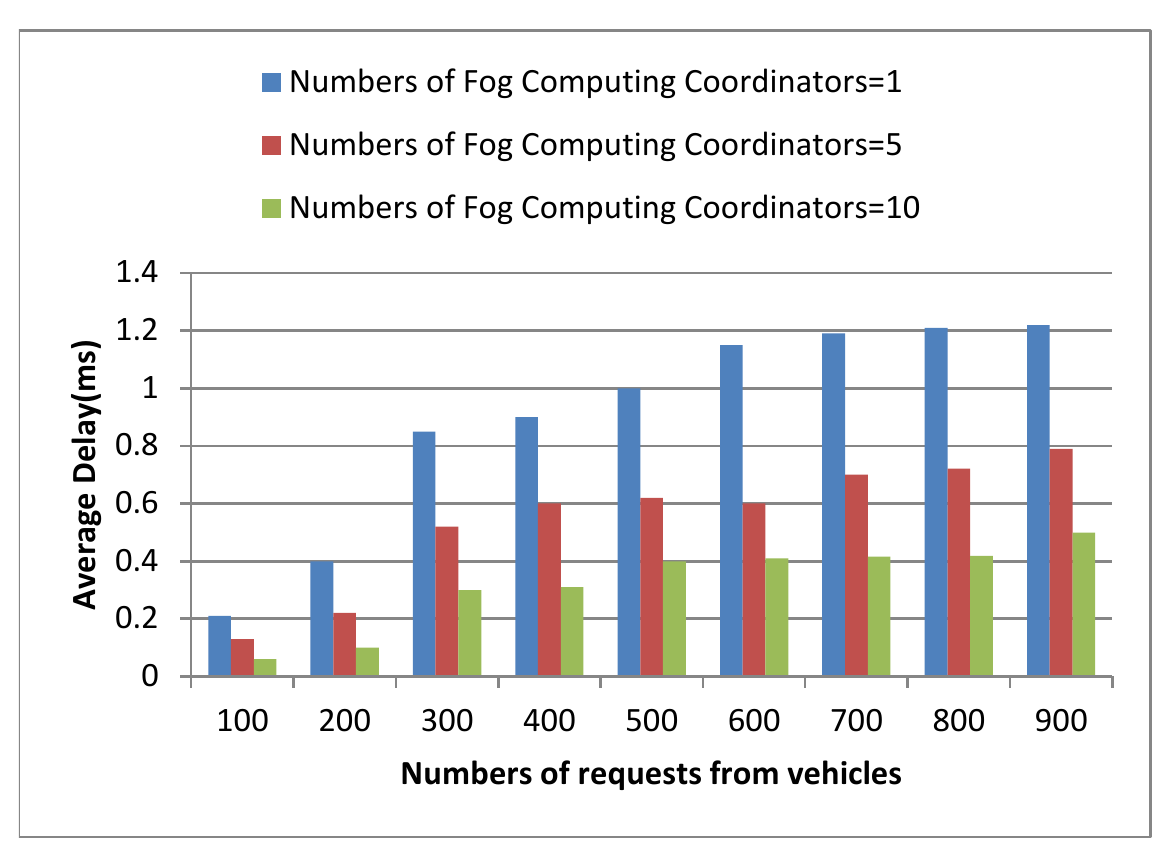} 
	\caption{Number of FNCs and application delay}\label{fig:Fig_6_FCN} 
\end{figure}

Fig.~\ref{fig:Fig_6_FCN} shows the relationship between the number of fog computing coordinators and application delay. As it shows, when the number of coordinators increases, the application delay decreases significantly. Therefore, the proposed architecture can effectively reduce the application delay of IoT applications in the smart grid.

\section{Conclusion}\label{sec:conclusion}

In this paper, we proposed a distributed computing architecture based on fog computing for IoT applications in the smart grid. We also proposed a programming model for the fog-based architecture. The major difference between our proposed architecture and the traditional one is the introduction of fog node coordination, which aims at better collaboration among fog nodes to meet various requirements in the smart grid. As demonstrated by evaluation and experimental results, our proposed architecture and programming model can significantly reduce service latency compared to the traditional fog-based architecture. In the future work, we will evaluate the system with a more practical communication protocol, and study the optimal resource allocation in this architecture. We will also consider nodes with high-speed mobility in more IoT applications.

\renewcommand\refname{Reference}
\bibliographystyle{IEEEtran}
\bibliography{IEEEfull,Reference}

\end{document}